\documentclass[aps,prl,showkeys,nofootinbib,superscriptaddress]{revtex4-2}

\usepackage[autostyle,italian=guillemets]{csquotes}
\usepackage[utf8]{inputenc}
\usepackage[nowrite,infront,standard]{frontespizio}
\usepackage[english]{babel}
\usepackage{physics}
\usepackage{amsmath,amssymb}
\usepackage{xcolor}
\usepackage{mathrsfs}
\usepackage{subfloat}
\usepackage{frontespizio}
\usepackage{chemmacros}
\usepackage{float}
\usepackage{graphicx} 
\usepackage{bm}
\usepackage{appendix}
\usepackage{hyperref}
\usepackage{midpage}
\usepackage{braket}
\usepackage{placeins}
\usepackage{subfig}
\usepackage{slashed}

\begin{document}
\title{Chiral oscillations in quantum field theory}
\author{Victor Bittencourt} 

\affiliation{ISIS (UMR 7006), Universit\'{e} de Strasbourg, 67000 Strasbourg, France}

\author{Massimo Blasone}
\affiliation{ Dipartimento di Fisica, Universit\`a degli Studi di Salerno, Via Giovanni Paolo II, 132 84084 Fisciano, Italy}
\affiliation{INFN, Sezione di Napoli, Gruppo Collegato di Salerno, Italy}

\author{Gennaro Zanfardino}

\affiliation{Dipartimento  di  Ingegneria Industriale, Universit\`a di Salerno, Via  Giovanni  Paolo  II, 132  I-84084  Fisciano  (SA),  Italy }
\affiliation{INFN, Sezione di Napoli, Gruppo Collegato di Salerno, Italy}

\date{\today}
%
\begin{abstract}
Dirac particles have two intrinsic degrees-of-freedom, helicity and chirality. While helicity is conserved in time, chirality is not constant under time evolution for massive particles, yielding the phenomenon of chiral oscillations. So far, chiral oscillations have been mainly described  in the framework of single particle relativistic quantum mechanics.
In this paper, we present a quantum field theory approach to chiral oscillations in analogy with the one used to describe flavor mixing and oscillations. 
By taking the expectation value of chiral charges, we obtain the same chiral oscillation formula derived via standard relativistic quantum mechanics. We find that chiral charges are diagonalized by a Bogoliubov transformation: this implies that the vacuum for particles with definite chirality is orthogonal to the one for those with definite energy.  In the case of neutrinos, our results can be further extended to include also flavor oscillations.
\end{abstract}

\maketitle

\section{Introduction}

Dirac particles have two intrinsic degrees-of-freedom: helicity and chirality \cite{WuTung}. While helicity is conserved during the time evolution, chirality is not. The mass term of the Dirac equation couples different chiral components of the bispinor: as such, a state with initially definite chirality will undergo the so-called chiral oscillations, whose frequency is proportional to the ratio between mass and energy \cite{Bernardini2005}. Charged weak interaction processes generate particles in states with definite chirality \cite{R_Pal,Pal:2010ih}, and hence such particles should exhibit chiral oscillations \cite{A01}. A particular implication is in the context of neutrino phenomenology, in which chiral oscillations modify the flavor oscillation formula \cite{Bernardini2005,Bernardini2006,Bernardini2011,Bittencourt2022}, affecting the detection of non-relativistic neutrinos, such as those from the cosmic neutrino background \cite{Bittencourt2021, Ge2020}.

So far, studies of chiral oscillations have mostly relied on a single-particle relativistic quantum mechanics approach \cite{Bernardini2005,Bernardini2006,Bernardini2011,Bittencourt2022, Esposito:1997, Esposito:1998, Bittencourt2021B,Bittencourt:2023brw,Bittencourt:2023iiu}, with some second quantization studies tailored for describing the interplay between flavor oscillations and chiral oscillations for neutrinos  \cite{Nishi2006}. 

In this work, we provide a description of chiral oscillations that follows the treatment of flavor mixing and oscillations in quantum field theory, introduced in Refs.\cite{bla95,bla98,Blasone:2001qa}. There, it was found that the vacuum state for the flavor fields is different from the one for the fields with definite masses. The connection between the two representations (flavor and mass) involves a Bogoliubov transformation. As a consequence of the particle-antiparticle condensate structure of the flavor vacuum, corrections to the standard oscillation formula \cite{Bilenky:1978nj,giunti2007fundamentals} appear in the nonrelativistic regime.

In analogy with this approach, we consider here the chiral charges for a Dirac field, in the presence of the  (chiral symmetry-breaking) mass term.  We show that these time-dependent charge operators are diagonalized by means of a Bogoliubov transformation. The chiral oscillation formula is then obtained as expectation value of the chiral charge and coincides with the one obtained within the first quantization approach. We also show that, for a massive Dirac field,   the vacuum state for particles with definite chirality  is orthogonal to the Dirac vacuum state.

\section{Chiral charges for a massive Dirac field}

Consider the  Dirac Lagrangian density
\begin{equation}
    \mathcal{L} \,= \,  \overline{\psi} \left( i \gamma^{\mu} \partial_{\mu} - m \right) \psi 
\end{equation}
The invariance of $\mathcal{L}$ under global phase transformations leads to the conserved charge
\begin{equation}
    Q = \int d^3{\bf x}\,\psi^{\dagger}(x)\psi(x)
\end{equation}

Dirac bispinor field $\psi$ can be splitted in left and chiral parts as $\psi = \psi_{L} + \psi_{R}$ where
\begin{equation}
    \psi_{L} \equiv P_L \psi(x) \, =\, \frac{1 - \gamma^{5}}{2} \psi(x), \qquad  \psi_{R} \equiv P_R \psi(x) \, =\,  \frac{1 + \gamma^{5}}{2} \psi(x).
\end{equation}
and hence Dirac Lagrangian can be written in terms of left and right chiral parts
\begin{equation}
    \mathcal{L}\, = \,\overline{\psi_{L}}   \,i\gamma^{\mu} \,\partial_{\mu}\, \psi_{L} \,+\,  \overline{\psi_{R}}  \,i\gamma^{\mu} \partial_{\mu} \psi_{R} - m\left(\overline{\psi_{L}}\,\psi_{R} \,+\, \overline{\psi_{R}}\,\psi_{L}\right)
\end{equation}
Because of the presence of the mass term, chiral symmetry is broken. 
Separate global phase transformations for $\psi_L$ and $\psi_R$ lead to the non-conserved chiral charges \cite{Cheng-Li}
\begin{equation}
    Q_{L}(t) = \int d^3{\bf x}\, \psi_{L}^\dagger (x)\psi_{L}(x), \qquad  Q_{R} (t)=  \int d^3{\bf x}\, \psi_{R}^\dagger(x) \psi_{R}(x).
\end{equation}
Note that the total (conserved) charge is equal to the sum of the (time dependent) chiral charges
\begin{equation}
    Q  \, = \,Q_{L}(t) \, +\, Q_{R}(t).
\end{equation}

\smallskip

We now proceed to quantize the fields $\psi$ and $\psi^{\dagger}$ by imposing the following (equal time) anticommutation relations
\begin{align}\label{CAR}
    &\{ \psi_{\alpha}(\textbf{x},t), \psi_{\beta}^{\dagger}(\textbf{y},t)\} = \delta_{\alpha, \beta} \delta^{3}(\textbf{x} - \textbf{y}) ,\nonumber \\[2mm]
    &\{ \psi_{\alpha}(\textbf{x},t), \psi_{\beta}(\textbf{y},t)\} = \{ \psi_{\alpha}^{\dagger}(\textbf{x},t), \psi_{\beta}^{\dagger}(\textbf{y},t)\} = 0
\end{align}
The expansion of the  field  in terms of creation and annihilation  is
\begin{equation}\psi(x) = \sum_{r=1,2}\int \frac{d^3{\bf k}}{(2\pi)^3} \Big[u^r_{\bf k} \,\alpha_{\bf k}^r\, e^{- i \omega_k t}\, +\, v^r_{-\bf k}\, \beta_{-\bf k}^{r\dag} \,e^{ i \omega_k t} \Big] e^{i {\bf k}\cdot{\bf x}}
\label{DF_number}
\end{equation}
with $\omega_k=\sqrt{k^2+m^2}$ and 
\begin{equation}
    \Big\{ \alpha_{\bf k}^r, \alpha_{\bf p}^{s\dag}\Big\} = \delta^3({\bf k}-{\bf p}) \delta_{rs}\, ;  \qquad
    \Big\{ \beta_{\bf k}^r, \beta_{\bf p}^{s\dag}\Big\} = \delta^3({\bf k}-{\bf p}) \delta_{rs}\, 
\end{equation}
and other anticommutators vanishing.
We denote by $|0\rangle$ the vacuum state annihilated by the above operators for the Dirac field:
\begin{equation}
     \alpha_{\bf k}^r\, |0\rangle \, = \, 
     \beta_{\bf k}^r\, |0\rangle \, = \, 0.
\end{equation}
Here $u$ and $v$ 
and satisfy the orthonormality and completeness relations:
\begin{eqnarray}
&&    u^{r\dag}_{\bf k}u^s_{\bf k}= \delta_{rs} \, , \quad 
 v^{r\dag}_{-\bf k}v^s_{-\bf k}= \delta_{rs}
 \, , \quad 
 u^{r\dag}_{-\bf k}v^s_{-\bf k}= 0
 \\
 &&
 \sum_r\left(u^r_{\bf k}u^{r\dag}_{\bf k} \,+\, v^r_{-\bf k}v^{r\dag}_{-\bf k} \right) \, = \, 1\!\!1
\end{eqnarray}

In terms of the creation and annihilation operators, the conserved charge $Q$ is written as
\begin{equation}
    Q = \sum_{r} \int d^3{\bf k} \left( \alpha_{\textbf{k}}^{r \dagger} \alpha_{\textbf{k}}^{r} - \beta_{-\textbf{k}}^{r \dagger} \beta_{-\textbf{k}}^{r} \right).
\end{equation}

If we now consider the chiral charges, we find:
\begin{eqnarray}
    Q_{L/R}(t) &= &\frac{1}{2} \left( Q \mp \int  d^3{\bf x}\, \psi^{\dagger}(x) \gamma^{5}\psi(x) \right)
\end{eqnarray}
Notice that the time dependence comes from the second piece in the above equation, which is indeed non-diagonal in the ladder operators:
\begin{eqnarray}
\nonumber
\int  d^3{\bf x}\, \psi^{\dagger}(x) \gamma^{5}\psi(x) &=& 
    \sum_{r=1,2} \int  d^3{\bf k}\,  \Bigg[(u^{r \dag}_{\bf k}  \,\gamma^{5}\, u^s_{\bf k})\,\alpha_{\bf k}^{r \dagger} \alpha_{\bf k}^r
\,+
\,\,(v^{r \dag}_{-\bf k}  \, \gamma^{5}\, v^s_{-\bf k}) \,\beta_{-{\bf k}}^r\beta_{-{\bf k}}^{r \dagger}\,
   \\ \label{offdiagterm}
&+&\, e^{- 2 i \omega_k t}\,
(v^{r \dag}_{-\bf k}  \, \gamma^{5}\, u^s_{\bf k}) \,
\beta_{-{\bf k}}^r\alpha_{\bf k}^r \,+\, e^{2 i \omega_k t}\,(u^{r \dag}_k  \,\gamma^{5} \, v^s_{- \bf k})\,\alpha_{\bf k}^{r \dagger}\beta_{-{\bf k}}^{r \dagger}  \Bigg].
\end{eqnarray}

By means of the following relations  (see Appendix) :
\begin{eqnarray}
\label{Bog1}
&& u^{r \dag}_{\bf k}  \, P_L\, u^s_{\bf k} \, = \,v^{r \dag}_{-\bf k}  \, P_R\, v^s_{-\bf k} \, = \, \frac{1}{2}\left(1+ \epsilon^r\frac{|{\bf k}|}{\omega_k}\right)\delta_{rs}
\\ \label{Bog2}
&& u^{r \dag}_{\bf k}  \, P_R\, u^s_{\bf k} \, = \,v^{r \dag}_{-\bf k}  \, P_L\, v^s_{-\bf k} \, = \, \frac{1}{2}\left(1- \epsilon^r\frac{|{\bf k}|}{\omega_k}\right)\delta_{rs}
\\ \label{Bog3}
&&u^{r \dag}_k  \, P_R\, v^s_{- \bf k} \, = \,- u^{r \dag}_k  \,P_L\, v^s_{- \bf k} \, = \, \frac{m}{2 \omega_k} \delta_{rs}
\end{eqnarray}
we obtain (after normal ordering):
\begin{eqnarray}
\label{Qleft}
{}\hspace{-3mm}    Q_{L}(t) &=& 
    \frac{1}{2}   \sum_{r=1,2} \int  d^3{\bf k}\,  \Bigg[\left(1+ \epsilon^r\frac{|{\bf k}|}{\omega_k}\right)\alpha_{\bf k}^{r \dagger} \alpha_{\bf k}^r
\,-
\,\left(1-\epsilon^r\frac{|{\bf k}|}{\omega_k}\right)\beta_{-{\bf k}}^{r \dagger} \beta_{-{\bf k}}^r\,
 -\, \frac{m}{\omega_k}\left( e^{- 2 i \omega_k t}\,\beta_{-{\bf k}}^r\alpha_{\bf k}^r + e^{2 i \omega_k t}\,\alpha_{\bf k}^{r \dagger}\beta_{-{\bf k}}^{r \dagger}\right)  \Bigg] \, ,
 \\
 \label{Qright}
  {}\hspace{-3mm} Q_{R}(t) &=& 
    \frac{1}{2}   \sum_{r=1,2} \int  d^3{\bf k}\,  \Bigg[\left(1- \epsilon^r\frac{|{\bf k}|}{\omega_k}\right)\alpha_{\bf k}^{r \dagger} \alpha_{\bf k}^r
\,-
\,\left(1+\epsilon^r\frac{|{\bf k}|}{\omega_k}\right)\beta_{-{\bf k}}^{r \dagger} \beta_{-{\bf k}}^r\,
 +\, \frac{m}{\omega_k}\left( e^{- 2 i \omega_k t}\,\beta_{-{\bf k}}^r\alpha_{\bf k}^r + e^{2 i \omega_k t}\,\alpha_{\bf k}^{r \dagger}\beta_{-{\bf k}}^{r \dagger}\right)  \Bigg] \, ,
\end{eqnarray}
with $\epsilon^r=(-1)^r$. In the relativistic limit $\omega_{k} \gg m$, the  off-diagonal time-dependent terms vanish and the chiral charges reduce to
\begin{eqnarray} \label{relchiralcharges1}
Q_L|_{m=0}&=&\int d^3{\bf k} \,\left(\alpha_{\bf k}^{2 \dagger} \alpha_{\bf k}^2
\,-
\,\beta_{-{\bf k}}^{1 \dagger} \beta_{-{\bf k}}^1\,\right)\, ,
    \\ \label{relchiralcharges2}
 Q_R|_{m=0}&=&\int d^3{\bf k} \,\left(\alpha_{\bf k}^{1 \dagger} \alpha_{\bf k}^1
\,-
\,\beta_{-{\bf k}}^{2 \dagger} \beta_{-{\bf k}}^2\,\right)\, ,
\end{eqnarray}
which are the conserved (Noether) charges for the Weyl fields $\psi_L$ and  $\psi_R$.

\section{Bogoliubov transformation and chiral vacuum structure}

We now consider the diagonalization of the charges \eqref{Qleft} and \eqref{Qright}.
To this aim, we  introduce the following canonical (Bogoliubov) transformation:
\begin{eqnarray}\label{aL}
    \alpha_{{\bf k},L} &=& \cos{\theta_k} \, \alpha_{\bf k}^2 \, - \, e^{i \phi_k}\, 
\sin{\theta_k}\,\beta_{-{\bf k}}^{2 \dagger}
    \\ \label{bL}
     \beta_{-{\bf k},L}^\dagger &=& \cos{\theta_k} \, \beta_{-{\bf k}}^{1 \dagger}  \, - \, e^{-i \phi_k}\, 
    \sin{\theta_k}\, \alpha_{{\bf k}}^1
    \\ \label{aR}
       \alpha_{{\bf k},R} &=& \cos{\theta_k} \, \alpha_{\bf k}^1 \, +  \, e^{i \phi_k}\, 
\sin{\theta_k}\,\beta_{-{\bf k}}^{1 \dagger}
    \\ \label{bR}
     \beta_{-{\bf k},R}^\dagger  &=& \cos{\theta_k} \, \beta_{-{\bf k}}^{2 \dagger} \,+  \, e^{-i \phi_k}\, 
    \sin{\theta_k}\, \alpha_{{\bf k}}^1
\end{eqnarray}
together with the inverse transformation
\begin{eqnarray}
    \alpha_{\bf k}^2 &=& \cos{\theta_k} \, \alpha_{{\bf k},L} \, + \,  \,e^{i \phi_k}\, 
    \sin{\theta_k}\, \beta_{-{\bf k},R}^{\dagger}
    \\
    \beta_{-{\bf k}}^{2\dagger} &=& \cos{\theta_k} \, \beta_{-{\bf k},R}^{ \dagger} \, - \,  \,e^{-i \phi_k}\, 
    \sin{\theta_k}\, \alpha_{{\bf k},L}
    \\
    \alpha_{\bf k}^1 &=& \cos{\theta_k} \, \alpha_{{\bf k},R} \, -  \,e^{i \phi_k}\, 
    \sin{\theta_k}\, \beta_{-{\bf k},L}^{\dagger}
    \\
    \beta_{-{\bf k}}^{1\dagger} &=& \cos{\theta_k} \, \beta_{-{\bf k},L}^{ \dagger} \, +   \,e^{-i \phi_k}\, 
    \sin{\theta_k}\, \alpha_{{\bf k},R}
\end{eqnarray}

By imposing diagonalization 
 of \eqref{Qleft} and \eqref{Qright} in terms of the new operators, and requiring that in the relativistic limit they 
reduce to \eqref{relchiralcharges1} and \eqref{relchiralcharges2}, we obtain the conditions\footnote{The coefficients of the Bogoliubov transformation are generated by the nontrivial products Eqs.\eqref{Bog1}-\eqref{Bog3}, in a similar way to the case of flavor mixing \cite{bla95} where they arise as products of spinors with different masses. }
\begin{eqnarray}
    \cos^{2}{\theta_k} &=& \frac{1}{2}\left( 1 + \frac{|{\bf k}|}{\omega_{k}}\right), \quad \sin^{2}{\theta_k} = \frac{1}{2}\left( 1 - \frac{|{\bf k}|}{\omega_{k}}\right), \nonumber \\ 
    \cos{2\theta_k} &=& \frac{|{\bf k}|}{\omega_{k}}, \quad 
    \sin{2\theta_k} = -\frac{m}{\omega_{k}}, \quad \phi_k = 2\omega_{k}t.
\label{parameter_transformation_2}
\end{eqnarray}
Thus the above defined chiral ladder operators are time-dependent and satisfy (equal time) canonical anticommutation relations (CAR):
\begin{equation}
    \Big\{ \alpha_{{\bf k},L}^r(t), \alpha_{{\bf k},L}^{s\dag}(t)\Big\} = \delta^3({\bf k}-{\bf p}) \delta_{rs}\, ,  \qquad
    \Big\{ \beta_{{\bf k},L}^r(t), \beta_{{\bf k},L}^{s\dag}(t)\Big\} = \delta^3({\bf k}-{\bf p}) \delta_{rs}\, 
\end{equation}

The diagonalization condition brings to the following form of the chiral charges in the presence of a mass term:
\begin{eqnarray}\label{QL}
    Q_L(t)&=&\int d^3{\bf k} \,\left(\alpha_{{\bf k},L}^{\dagger}(t)\alpha_{{\bf k},L}(t)
    - \,\beta_{-{\bf k},L}^{ \dagger}(t)\beta_{-{\bf k},L}(t) \right), 
    \\ \label{QR}
    Q_R(t)&=&\int d^3{\bf k} \,\left(\alpha_{{\bf k},R}^{ \dagger}(t)\alpha_{{\bf k},R}(t)
    - \,\beta_{-{\bf k},R}^{ \dagger}(t)\beta_{-{\bf k},R}(t) \right), 
\end{eqnarray}
consistent with the boundary conditions \eqref{relchiralcharges1}, \eqref{relchiralcharges2} in the relativistic limit $\frac{m}{\omega_{k}} \to 0$. 

Note that in Eqs.\eqref{QL},\eqref{QR} the the sum on the spins present in Eqs.\eqref{Qleft}, \eqref{Qright} disappears.
This happens because the  field $\psi_{L}$, $\psi_{R}$ carries each only ``one half"  of the DoFs of the Dirac field.

\vspace{0.2cm}
Furthermore, it is possible to expand the Dirac field in term of the chiral creation/annihilation operators:
\begin{eqnarray}\nonumber
\psi(x) &=& \int \frac{d^3{\bf k}}{(2\pi)^3} e^{i {\bf k}\cdot{\bf x}}\Bigg[u^1_{\bf k}\left(\cos{\theta_k} \, \alpha_{{\bf k},R} \, - \,  \,e^{i \phi_k}\, 
    \sin{\theta_k}\, \beta_{-{\bf k},L}^{\dagger}
\right) e^{- i \omega_k t}
\\ \nonumber
&&+u^2_{\bf k}\left(\cos{\theta_k} \, \alpha_{{\bf k},L} \, +  \,e^{i \phi_k}\, 
    \sin{\theta_k}\, \beta_{-{\bf k},R}^{\dagger}
\right) e^{- i \omega_k t}
\\ \nonumber
&&+v^1_{-\bf k}\, 
\left(\cos{\theta_k} \, \beta_{-{\bf k},L}^{ \dagger} \, + \,  \,e^{-i \phi_k}\, 
    \sin{\theta_k}\, \alpha_{{\bf k},R}
\right)
\,e^{ i \omega_k t}
\\ 
&&+v^2_{-\bf k}\, 
\left(\cos{\theta_k} \, \beta_{-{\bf k},R}^{ \dagger} \, -   \,e^{-i \phi_k}\, 
    \sin{\theta_k}\, \alpha_{{\bf k},L}
\right)
\,e^{ i \omega_k t}\Bigg] ,
\end{eqnarray}
which can be rearranged in the following form (using $\phi_k=2 \omega_k t$):
\begin{eqnarray} \nonumber
\psi(x) &=& \int \frac{d^3{\bf k}}{(2\pi)^3} \Big[u_{{\bf k},L} \,\alpha_{{\bf k},L}(t)\,e^{ -i \omega_k t}\, +\, v_{{-\bf k},L}\, \beta_{{-\bf k},L}^{\dag}(t)\,e^{ i \omega_k t}  \Big] e^{i {\bf k}\cdot{\bf x}}
\\  \nonumber
&+& \int \frac{d^3{\bf k}}{(2\pi)^3} \Big[u_{{\bf k},R} \,\alpha_{{\bf k},R}(t)\,e^{ -i \omega_k t}\, +\, v_{{-\bf k},R}\, \beta_{{-\bf k},R}^{\dag}(t)\,e^{ i \omega_k t}  \Big] e^{i {\bf k}\cdot{\bf x}}
\\  [2mm]
&=& \psi_L(x) \, + \,  \psi_R(x)
\end{eqnarray}
with
\begin{eqnarray}
    u_{{\bf k},L} &\equiv& \cos{\theta_k} \, u^2_{\bf k} \, - \, \sin{\theta_k} \, v^2_{-\bf k}, \quad 
    u_{{\bf k},R} \equiv  \cos{\theta_k} \, u^1_{\bf k} \, + \, \sin{\theta_k} \, v^1_{-\bf k} \\ [2mm]
    v_{-{\bf k},L} &\equiv& \cos{\theta_k} \, v^1_{-{\bf k}} \, - \, \sin{\theta_k} \, u^1_{\bf k} ,\quad
    v_{-{\bf k},R} \equiv \cos{\theta_k} \, v^2_{-{\bf k}} \, + \, \sin{\theta_k} \, u^2_{\bf k}
\end{eqnarray}

It is easy to verify that the above defined spinors satify the 
orthonormality relations:
\begin{eqnarray}
&& u^{\dag}_{{\bf k},L}u_{{\bf k},L}\,=\, 
u^{\dag}_{{\bf k},R}u_{{\bf k},R}\,=\, 1\, , \quad 
  v^{\dag}_{-{\bf k},L}v_{-{\bf k},L}\,=\,v^{\dag}_{-{\bf k},R}v_{-{\bf k},R}\,=\, 1
\\[1mm]
&&
u^{\dag}_{{\bf k},L}u_{{\bf k},R}\,=\, 
v^{\dag}_{-{\bf k},L}v_{-{\bf k},R}\,=\, 0\, , \quad 
  u^{\dag}_{{\bf k},L}v_{-{\bf k},L}\,=\,u^{\dag}_{{\bf k},R}v_{-{\bf k},L}\,=\,0
\end{eqnarray}
and the completeness relation:
\begin{equation}
    u_{{\bf k},R}\,u^{\dag}_{{\bf k},R} \, +\, 
 u_{{\bf k},L}\,u^{\dag}_{{\bf k},L} \, +\, v_{-{\bf k},R}\,v^{\dag}_{-{\bf k},R}
 +\, v_{-{\bf k},L}\,v^{\dag}_{-{\bf k},L} \, =\,  1\!\!1
\end{equation}

They also fulfill the following consistency relations:
 \begin{eqnarray} 
   P_L\, u_{{\bf k},L} &=&
   \, u_{{\bf k},L}
   \,,\qquad 
   P_L\, v_{-{\bf k},L} \,=
   \, v_{-{\bf k},L} 
   \\ 
   P_R\, u_{{\bf k},R} &=&
   \, u_{{\bf k},R}
   \,,\qquad 
   P_R\, v_{-{\bf k},R} \,=
   \, v_{-{\bf k},R} 
   \\ 
    P_R\, u_{{\bf k},L} &=&  P_R\, v_{-{\bf k},L}\,=
   \,  P_L\, u_{{\bf k},R}\,=
   \,  P_L\, v_{-{\bf k},R}\,=
   \, 0.
 \end{eqnarray}

Finally, we study the (time dependent)  vacuum state for the operators \eqref{aL}-\eqref{bR}. The above (Bogoliubov) transformation can be recast in the form
\begin{eqnarray}
    \alpha_{{\bf k},L} (t)&=&
    G^{-1}_t \alpha_{\bf k}^2 G_t \qquad, \quad \beta_{{\bf k},L}(t) \,=\,
    G^{-1}_t \beta_{\bf k}^1 G_t
    \\ [2mm]
     \alpha_{{\bf k},R}(t) &=&
    G^{-1}_t \,\alpha_{\bf k}^1 \,G_t\qquad, \quad \beta_{{\bf k},R}(t) \,=\,
    G^{-1}_t \,\beta_{\bf k}^2 \,G_t
\end{eqnarray}
with the generator $G_t$ given by\footnote{Notice the time dependence of the generator, due to $\phi_k(t)$.}
\begin{equation}
    G_t(\theta,\phi)\, = \, 
    \exp \left[\sum_r\int d^3{\bf k} \, \theta_k \epsilon^r \left( e^{- i \phi_k(t)}\alpha_{\bf k}^r \beta_{-{\bf k}}^r - 
    e^{ i \phi_k(t)}\beta_{-{\bf k}}^{r \dag} \alpha_{\bf k}^{r \dag} \right)\right]
\end{equation}
where we made explicit the time dependence of $\phi_k$.
The explicit form for the \emph{chiral vacuum} for a massive field at time $t$ is
\begin{equation}
    |{\tilde 0} (t) \rangle_{LR} \, = \, \prod_{{\bf k},r}\left[\cos{\theta_k} \,+\, \epsilon^r e^{ i \phi_k(t)} \sin{\theta_k} \, \alpha_{\bf k}^{r \dag}\beta_{-{\bf k}}^{r \dag}\right]| 0\rangle
\end{equation}

In a similar way as done in Ref.\cite{bla95}, it is possible to prove that the chiral vacuum $|{\tilde 0} (t) \rangle_{LR}$ and the Dirac vacuum $|0\rangle$ are orthogonal in the infinite volume limit:
\begin{equation}
    \lim_{V\rightarrow\infty} \,  \langle 0|{\tilde 0} (t) \rangle_{LR} \, = \, 0, 
\end{equation}
thus generating unitarily inequivalent representations of the field algebra \eqref{CAR}.

\section{chiral oscillation formula}

At this point it is possible to compute the expectation value of the left-charge $Q_{L}$ on a state with a well defined chirality. Hence we define the state $|\alpha_L\rangle\equiv\alpha_{{\bf k},L}^{ \dagger}|{\tilde 0} \rangle_{LR} $,
with $|{\tilde 0} \rangle_{LR} \equiv |{\tilde 0} (0) \rangle_{LR}$ is the vacuum for the chiral excitations at reference time $t=0$. An analogous calculation can be done for an initial right chiral state.

Considering that $\phi_k=2 \omega_{k}t$, we have that the left chiral operator at time $t$ is
\begin{equation}
    \alpha_{{\bf k},L}(t) = \cos{\theta_k} \, e^{-  i \omega_k t}\,\alpha_{\bf k}^2 \, -  \, 
    \sin{\theta_k}\,e^{ i \omega_k t}\,\beta_{-{\bf k}}^{2 \dagger}.
\end{equation}
The oscillation formula is therefore:
\begin{equation}
    \langle \alpha_{{\bf k},L} | Q_L(t) |\alpha_{{\bf k},L} \rangle
= |\{\alpha_{{\bf k},L}(t),\alpha_{{\bf k},L}^{\dagger}(0)\}|^2 
\end{equation}
with
\begin{equation}
    \{\alpha_{{\bf k},L}(t),\alpha_{{\bf k},L}^{\dagger}(0)\} = \cos^2\theta_k e^{-  i \omega_k t} + \sin^2\theta
_k e^{ i \omega_k t}
\end{equation}
We thus obtain
\begin{equation}\label{chirosc}
    \langle \alpha_{{\bf k},L} | Q_L(t) |\alpha_{{\bf k},L} \rangle
= 1 - \sin^2(2\theta_k) \sin^2(\omega_k t)
= 1 - \frac{m^2}{\omega_k^2} \sin^2(\omega_k t)
\end{equation}
This result coincides with the oscillation formula obtained independently in \cite{A01,Bittencourt2021}.

For convenience of the reader, we report here the explicit calculation for the above  formula:
\begin{eqnarray} \nonumber
    &&{}\hspace{-15mm}\langle \alpha_{{\bf k},L} | \Big( \alpha_{{\bf k},L}^{ \dagger}(t)\alpha_{{\bf k},L}(t) - \beta_{-{\bf k},L}^{ \dagger}(t)\beta_{-{\bf k},L}(t)\Big)|\alpha_{{\bf k},L} \rangle =
     |\{\alpha_{{\bf k},L}(t),\alpha_{{\bf k},L}^{\dagger}(0)\}|^2 
     \\ 
     &&{}\hspace{-15mm} + |\{\alpha_{{\bf k},L}(t),\beta_{-{\bf k},L}(0)\}|^2  
    + {}_{LR}\langle{\tilde 0}| \alpha_{{\bf k},L}^{ \dagger}(t)\alpha_{{\bf k},L}(t)|{\tilde 0} \rangle_{LR} - 
    {}_{LR}\langle{\tilde 0}| \beta_{-{\bf k},L}^{ \dagger}(t)\beta_{-{\bf k},L}(t)|{\tilde 0} \rangle_{LR}\, ,
\end{eqnarray}
where 
\begin{eqnarray}
&& {}\hspace{-10mm}\{\alpha_{{\bf k},L}(t),\beta_{-{\bf k},L}(0)\}=\cos{\theta_k}\sin{\theta_k} \left( e^{-  i \omega_k t} e^{i \phi}\,- \, e^{  i \omega_k t}\right) \,=\,0 \,,
\\
&&{}\hspace{-10mm}
{}_{LR}\langle{\tilde 0}| \alpha_{{\bf k},L}^{ \dagger}(t)\alpha_{{\bf k},L}(t)|{\tilde 0} \rangle_{LR} = 
    {}_{LR}\langle{\tilde 0}| \beta_{-{\bf k},L}^{ \dagger}(t)\beta_{-{\bf k},L}(t)|{\tilde 0} \rangle_{LR}=\frac{m^2}{\omega_k^2} \sin^2(\omega_k t)\, .
\end{eqnarray}

\section{Conclusions}

In this work we have analyzed  chiral oscillations in a quantum field theory framework. We have adopted the same approach used for describing flavor mixing and oscillations \cite{bla95,bla98,Blasone:2001qa}, thus extending previous studies to chiral oscillations based on Dirac equation \cite{Bernardini2005,Bernardini2006,Bernardini2011,Bittencourt2022}.  By taking the expectation value of the (time-dependent) chiral charge operator on a state with definite chirality, we have derived the same chiral oscillation formula obtained in relativistic quantum mechanics.

Moreover, we have shown that the relation between the creation/annihilation operators for particles with definite chirality with those with definite energy is indeed a Bogoliubov transformation.
The expression for the chiral vacuum for massive fields is then that of a condensate of particle-antiparticle pairs, similar to the BCS ground state.
The chiral vacuum and the Dirac vacuum turn out to be orthogonal in the infinite volume limit, thus giving rise to unitarily inequivalent representations of the field algebra. 

This result appears to be quite fundamental and of similar nature of what found in Ref.\cite{bla95}, where it was proved the unitary inequivalence of the mass and the flavor representations. In both cases, it appears that  weak interactions act in a non-trivial way at the level of representation, where particle states are defined. 

Various aspects and phenomenological consequences of the results obtained in this paper  are under consideration and will be object of future publications. These include the issue of Lorentz (Poincar\'e) invariance of the chiral vacuum, time-energy uncertainty relations, contribution to  the vacuum energy from the condensate structure, etc.

Finally, following an approach recently developed in Refs.\cite{Blasone:2023brf}, where the QFT flavor oscillation formula was (approximately) recovered in a perturbative calculation at finite time, a similar treatment for chiral oscillations is being carried out \cite{ChiralInt}.

\appendix

\section{Appendix: useful relations}

Spinor wavefunctions and Dirac matrices:
\begin{equation}
    u_{\bf k}^{r}=A_k\;
\left(\begin{array}{c} \xi^r 
\\  [2mm] \frac{{\bf \sigma}\cdot{\bf k}}
{\omega_{k}+m} 
\xi^r\end{array}\right) \, ; \qquad
v_{-{\bf k}}^{r}=A_k\;
\left(\begin{array}{c}\frac{-{\bf \sigma}\cdot{\bf k}}
{\omega_{k}+m} 
\xi^r \\ [2mm] \xi^r \end{array}\right)
\end{equation}
with $A_k= \left(\frac{\omega_{k}+m}{2\omega_{k}}\right)
^{\frac{1}{2}}$ and $\xi^1=\left(\begin{array}{c} 1 
\\  [2mm]0\end{array}\right)$ and 
$\xi^2=\left(\begin{array}{c} 0 
\\  [2mm]1\end{array}\right)$.
We also have
$
{\bf \sigma}\cdot{\bf k}=\left(\begin{array}{cc} k_{3} & k_{1}-ik_{2} \\ 
k_{1}+ik_{2} & -k_{3}\end{array}
\right)$
and
\begin{equation}
    \gamma_{i}=\left(\begin{array}{cc} 0 & \sigma_{i} \\ - \sigma_{i} & 0
\end{array}\right),\qquad\quad
\gamma_{0}=\left(\begin{array}{cc} 1_{2} & 0 \\ 0 & -1_{2}
\end{array}\right),\qquad\quad
\gamma_5 =\left(\begin{array}{cc} 0 & 1_{2}  \\ 1_{2}& 0 
\end{array}\right)
\end{equation}

Calculation of coefficients appearing in the off--diagonal term \eqref{offdiagterm}:
\begin{eqnarray}
&& u^{1 \dag}_{\bf k}  \gamma_5 u^1_{\bf k} = \frac{|{\bf k}|}{\omega_k}, \qquad u^{2 \dag}_{\bf k}  \gamma_5 u^2_{\bf k} = -\frac{|{\bf k}|}{\omega_k}, \qquad u^{1 \dag}_{\bf k}  \gamma_5 u^2_{\bf k} =0
\\
&& u^{1 \dag}_{\bf k}  \gamma_5 v^1_{-{\bf k}} =  u^{2 \dag}_{\bf k}  \gamma_5 v^2_{-{\bf k}} =\frac{m}{\omega_k}, \qquad u^{1 \dag}_{\bf k}  \gamma_5 v^2_{-{\bf k}} = 0
\\ 
&& v^{1 \dag}_{-\bf k}  \gamma_5 v^1_{-\bf k} = -\frac{|{\bf k}|}{\omega_k}, \qquad
v^{2 \dag}_{-\bf k}  \gamma_5 v^2_{-\bf k} = \frac{|{\bf k}|}{\omega_k}
\end{eqnarray}

The above relations are easily checked in the special reference frame ${\bf k}=(0,0,|{\bf k}|)$, where the spinors assume the simple form
%
%
%
\begin{equation}
u_{\bf k}^{1}=
A_k\;
\left(\begin{array}{c} 1 \\ 0 \\ \frac{|{\bf k}|}{\omega_{k}+m } \\
 0\end{array}\right)
\; \;\;\;\;\;
u_{\bf k}^{2}=
A_k\;
\left(\begin{array}{c} 0 \\ 1 \\ 0  \\
 \frac{-|{\bf k}|}{\omega_{k}+m}\end{array}\right)
 \; \;\;\;\;\;
 v_{-\bf k}^{1}=
A_k\;
\left(\begin{array}{c}  \frac{-|{\bf k}|}{\omega_{k}+m} \\
 0 \\ 1 \\ 0\end{array}\right)
\; \;\;\;\;\;
v_{-\bf k}^{2}=
A_k\;
\left(\begin{array}{c}  0 \\
 \frac{|{\bf k}|}{\omega_{k}+m} \\ 0 \\ 1\end{array}\right)
\end{equation}

\end{document}